\documentstyle[eqsecnum,aps,prd]{revtex}
\begin{document}
\draft
\thispagestyle{empty}
{\baselineskip0pt
\leftline{\large\baselineskip16pt\sl\vbox to0pt{\hbox{Department of Physics}
               \hbox{Kyoto University}\vss}}
\rightline{\large\baselineskip16pt\rm\vbox to20pt{\hbox{KUNS 1305}
           \hbox{Dec. 1994}
\vss}}%
}
\vskip3cm
\begin{center}{\large
Particle Spectrum Created through Bubble Nucleation \\
and \\
Quantum Field Theory in the Milne Universe}
\end{center}
\begin{center}
 {\large Kazuhiro Yamamoto, Takahiro Tanaka and Misao Sasaki} \\
{\em Department of Physics,~Kyoto University} \\
{\em Kyoto 606-01,~Japan}
\end{center}
\begin{abstract}
Using the multi-dimensional wave function formalism, we investigate
the quantum state of a scalar field inside a true vacuum bubble
nucleated through false vacuum decay in flat spacetime.
We developed a formalism which allows us a mode-by-mode analysis.
To demonstrate its advantage, we describe in detail the evolution
of the quantum state during the tunneling process in terms of
individual mode functions and interpret the result in the language
of particle creation.
The spectrum of the created particles is examined based on quantum
field theory in the Milne universe.


\end{abstract}


\section{Introduction}

{} Various phenomena associated with phase transitions at the early stage of
the universe have been a subject of great interest in cosmology
for two decades. Cosmological first order phase transitions are
caused by the decay of a metastable (false) vacuum in field theory,
which produces true vacuum bubbles in the sea of false vacuum.
A particular interest was taken in the false vacuum decay during
an inflationary stage of the universe
\cite{Ref:Guth,Ref:Sato,Ref:LaStein}.

Recently, several authors considered a possible inflationary
universe scenario which realize $\Omega_0\sim0.1$ open universe
\cite{Ref:Gott,Ref:Recent,Ref:Openinf},
in contrast to the standard inflationary universe which predicts
$\Omega_0=1$ flat universe \cite{Ref:KTF}.
In this scenario, the bubble nucleation process plays an important role.
That is, one assumes a scalar field with a potential
like in the new inflationary scenario but with a high potential barrier
before the slow rolling inflationary phase.
If the nucleation rate is small, nucleated bubbles do not
collide, and a homogeneous and isotropic universe with negative
curvature is realized in one bubble thanks to the $O(4)$-symmetry
of the bounce solution.
The second inflation starts inside a bubble due to the flat part
of the potential. Then after the inflation, the universe gradually
becomes curvature-dominated as it expands and the present universe with
$\Omega_0\sim0.1$ can be explained.
The important problem is the quantum state of the scalar field after
bubble nucleation because its quantum fluctuations may be
considered as the origin of the cosmic structure.

As is well-known, the false vacuum decay rate can
be calculated by means of the Euclidean path integral dominated
by an $O(4)$-symmetric bounce solution and the classical motion of the
nucleated bubble is described by the analytic continuation of the bounce
solution \cite{Ref:Coleman,Ref:CC}.
But studies of the quantum state after the false vacuum decay have
not been done so much.
Several pioneering works on this problem were done by
Rubakov \cite{Ref:Rubakov}, Kandrup \cite{Ref:Kandrup}, Vachaspati
and Vilenkin \cite{Ref:VV}.
Recently, we developed a formalism to investigate the quantum
state after false vacuum decay based on the WKB
wave function in a multi-dimensional tunneling system
\cite{Ref:TSY,Ref:STYYa,Ref:STYYb,Ref:TS},
which was originally developed by Banks, Bender and Wu \cite{Ref:BBW},
and Gervais and Sakita \cite{Ref:GS}.

For a simplified model of spatially homogeneous false vacuum decay
discussed by Rubakov \cite{Ref:Rubakov},
we showed that our formalism leads to the same
results \cite{Ref:TSY}.
As for the false vacuum decay due to an $O(4)$-symmetric
vacuum bubble, we also applied our formalism to a simple toy model
and evaluated the expectation value of the energy momentum tensor
inside the vacuum bubble \cite{Ref:STYYa}.
The result indicates that the field excitation
occurs inside the vacuum bubble, {\it i.e.}, particle creation occurs
there and it resembles a thermal state.

In spite of these investigations, we have not yet obtained a clear
understanding of the quantum state after bubble nucleation.
Neither the spectrum of created particles, nor  the relation
between the efficiency of the particle creation and the shape of
the field potential has not been made clear.
A difficulty originates from the fact that the mass of the
quantum field is spatially inhomogeneous and time-dependent due to
the non-trivial background field configuration.

In this paper, we tackle the issue by looking at the evolution of
mode functions of a scalar field in a systematic manner
and by investigating the spectrum of created particles,
to obtain a better understanding of the quantum
state after bubble nucleation.
Our strategy is as follows.
The background field configuration is inhomogeneous and time-dependent
in the conventional Minkowski space-time coordinates.
However, it keeps the Lorentz group ($O(3,1)$) symmetry, which originates from
the $O(4)$-symmetry of the bounce solution.
This implies that the system is homogeneous on the hyperbolic time
slicing inside the future light cone within one bubble.
This time slicing of the Minkowski spacetime is called the Milne
universe.
Quantum field theory in the Milne universe has been well
investigated as an example of the quantization of a field in
curved spacetime \cite{Ref:BD,Ref:diSessa,Ref:Somm,Ref:Gromes},
and the relation of the vacuum state between the Milne universe
and the Minkowski spacetime is well understood for a massive field, at
least for two-dimensional spacetime.
Using this fact, we can define a natural vacuum state in the Milne
universe that corresponds to the usual Minkowski vacuum.
Then we can gain an insight into the quantum state of a field inside an
 expanding bubble in terms of the particle spectrum observed by the
comoving observers in the Milne universe at late times.

Following this strategy, we first
carefully investigate the evolution of the mode functions through the
tunneling process in two-dimensional spacetime, based on the wave functional
formalism recently developed by us \cite{Ref:TSY}.
We find that the resulting mode functions can be expressed in a
very transparent manner and the created particle spectrum can be evaluated
mode by mode. Then we extend the result to the case of four dimensional
spacetime. This technical advantage will be very useful
when we consider realistic cases of bubble nucleation.

The paper is organized as follows.
In section 2, we review our formalism to investigate the quantum state
after bubble nucleation.
In section 3, we investigate the quantum state inside a vacuum bubble in
two-dimensional spacetime on the basis of quantum field theory in the
Milne universe, and derive the formula for the spectrum of
created particles.
Then the result is extended to the four-dimensional case in section 4.
In section 5, we apply our result to a simple thin-wall model
which is similar to the one discussed by Rubakov \cite{Ref:Rubakov}.
The results are found to be similar to those obtained by Rubakov
in the case of a spatially homogeneous decay model.
Section 6 is devoted to summary and discussions.
We use the units $\hbar=1$ and $c=1$ and a bar ($\bar{\ }$)
to denote the complex conjugate.

\section{Review of Formalism}
First, let us start reviewing our formalism to investigate the
quantum state after false vacuum decay \cite{Ref:TSY}.
 It is based on the
WKB wave function for a multi-dimensional tunneling system.
For generality, we consider the $(n+1)$-dimensional Minkowski spacetime.
The conventional Minkowski spacetime coordinates are denoted by
$(t, {\bf x})$ with ${\bf x}=(x_1,\cdots,x_n)$.
We consider a system of two interacting real scalar fields
 $\sigma$ and $\phi$ with the
Lagrangian \cite{Ref:Rubakov},
\begin{equation}
{\cal L}={\cal L}_\sigma+{\cal L}_\phi,
\end{equation}
where
\begin{eqnarray}
{\cal L}_{\sigma}&=&-\int d^n{\bf x}
\Biggl[{1\over2} (\partial_\mu\sigma)^2
+V(\sigma)\Biggr],
\\
{\cal L}_{\phi}&=&-\int d^n{\bf x}
\Biggl[{1\over2} (\partial_\mu\phi)^2
+{1\over2}m^2(\sigma)\phi^2\Biggr],
\end{eqnarray}
Here the $\sigma$ field has a potential $V(\sigma)$ as shown in
Figure 1, and decays from the false vacuum at $\sigma({\bf x})=\sigma_F$
to the true vacuum at $\sigma({\bf x})=\sigma_T$ through bubble
nucleation.

In the lowest WKB picture, the decay from the false vacuum
can be described by a bounce solution \cite{Ref:Coleman}.
The bounce solution
satisfies the classical Euclidean field equation, and has the
maximum symmetry to minimize the Euclidean action.
In our case, it has the $O(n+1)$-symmetry,
and depends only on $\tau^2+{\bf x}^2$, where $\tau$ ($=it$)
is the time coordinate in the Euclidean space.
Thus we can write the bounce solution
as $\sigma_0 (\tau^2+{\bf x}^2)$.
The subsequent motion of the bubble after nucleation is given
by the analytic continuation of the bounce
solution to the Lorentzian time, which is given by ${\sigma_0}(-t^2+{\bf
x}^2)$.

For simplicity, we neglect the quantum fluctuations
of the $\sigma$ field and consider only the quantum state of
the $\phi$ field which will be affected through the coupling
with $\sigma$ as it decays from the false vacuum.
Note that one may regard $\phi$ as the fluctuation of $\sigma$ itself,
if desired \cite{Ref:Hama}.
The wave function that describes the quantum state in the Euclidean
region is written as \cite{Ref:TSY,Ref:STYYa}
\begin{equation}
{\Psi}_E={\cal N}(\tau) e^{-S_E[{\sigma_0}]} {\psi}_E[\tau,\phi(\cdot)],
\label{eq:EWF}
\end{equation}
where
\begin{eqnarray}
&&S_E[{\sigma_0}]=\int_{-\infty}^{\tau} d\tau'
 \int d^n{\bf x}\Biggl[
{1\over2}\biggl({\partial {\sigma_0}\over \partial\tau'}\biggr)^{2}
+{1\over2}\biggl({\partial {\sigma_0}\over \partial {\bf x}}\biggr)^{2}
+V({\sigma_0}) \Biggr],
\\
&& {\psi}_E[\tau,\phi(\cdot)]=
\exp\Biggl[
-{1\over2}\int\int d^n{\bf x} d^n{\bf y} \phi({\bf x})
\Omega_E({\bf x},{\bf y};\tau)\phi({\bf y})\Biggr].
\end{eqnarray}
In Eq.(\ref{eq:EWF}), $e^{-S_E[{\sigma_0}]}$ is the
lowest WKB part and gives the classical picture of the tunneling
described by the bounce solution $\sigma_0$.
The factor ${\cal N}(\tau)$ describes
the next WKB order corrections to the normalization factor
and is independent of $\phi$.
The part ${\psi}_E$ is the wave functional that describes the
state of the field $\phi$ on the classical background of $\sigma_0$.
Thus all the information of the quantum state is contained in the
function $\Omega_E({\bf x},{\bf y};\tau)$. It is expressed in terms of
the Euclidean mode functions ${g_{{\bf k}}(\tau,{\bf x})}$ as
\begin{equation}
\Omega_E({\bf x},{\bf y};\tau):=\int d^n{\bf k} {\partial {g_{{\bf
k}}(\tau,{\bf x})}\over \partial \tau}
g_{\bf k}^{-1}(\tau,{\bf y}),
\end{equation}
where ${g_{{\bf k}}(\tau,{\bf x})}$ satisfies the Euclidean field equation,
\begin{equation}
\left[{\partial^2\over\partial \tau^2}
     +\sum_{i=1}^n{\partial^2\over\partial x_i^2}
     -m^2\big({\sigma_0}(\tau^2+{\bf x}^2)\big)
\right]{g_{{\bf k}}(\tau,{\bf x})}=0,
\label{eq:Eqofg}
\end{equation}
with the boundary condition at $\tau\rightarrow-\infty$ as
\begin{equation}
g_{\bf k}\rightarrow e^{\omega_k\tau} Y_{\bf k}({\bf x}),
\label{eq:boundary}
\end{equation}
where $\omega_k:=\sqrt{{\bf k}^2+m^2\big(\sigma_F\big)}$,
$\sigma_F={\sigma_0}(\tau\rightarrow-\infty)$, and $Y_{\bf k}({\bf x})$ is a
spatial
harmonic function.
This boundary condition is fixed by requiring that the system
is set initially in a ground state in the false vacuum.
The inverse of ${g_{{\bf k}}(\tau,{\bf x})}$, $g_{\bf k}(\tau,{\bf x})^{-1}$ is
defined by
\begin{equation}
\int d^n{\bf x} g_{\bf k}(\tau,{\bf x}) g_{{\bf k}'}(\tau,{\bf
x})^{-1}=\delta({\bf k}-{\bf k}').
\end{equation}

The quantum state after false vacuum decay is described by
the analytic continuation of the above wave functional into the
Lorentzian region ($\tau\rightarrow it$)
where the classical motion is allowed \cite{Ref:TSY}:
\begin{equation}
{\Psi}_L\propto e^{iS[{\sigma_0}]} {\psi}_L[t,\phi(\cdot)],
\label{eq:LWF}
\end{equation}
where
\begin{eqnarray}
&&
S[{\sigma_0}]=\int_{0}^{t} dt'{\cal L}_\sigma[\sigma_0(-t'^2+{\bf x}^2)],
\\
&&{\psi}_L[t,\phi(\cdot)]=\exp\Biggl[
-{1\over2}\int\int d^n{\bf x} d^n{\bf y} \phi({\bf x})
\Omega({\bf x},{\bf y};t)\phi({\bf y})\Biggr],
\end{eqnarray}
and $\Omega({\bf x},{\bf y};t)$ is expressed in terms of the
mode functions $\overline{{u_{{\bf k}}(t,{\bf x})}}$, which is the analytic
continuation of
${g_{{\bf k}}(\tau,{\bf x})}$ to the Lorentzian time, as
\begin{equation}
\Omega({\bf x},{\bf y};t):={1\over i}\int d^n{\bf k}
{\partial \overline{{u_{{\bf k}}(t,{\bf x})}}\over \partial t}
\overline{u^{-1}_{\bf k}(t,{\bf y})}.
\end{equation}
Just as in the case of Eq.(\ref{eq:EWF}),
the first part $e^{iS[{\sigma_0}]}$ in Eq.(\ref{eq:LWF}) describes the
classical
 motion of the expanding bubble, and the second part ${\psi}_L[t,\phi(\cdot)]$
does the quantum state of the $\phi$ field on the classical
background of the expanding bubble ${\sigma_0}({\bf x}^2-t^2)$.

It has been shown that the quantum state of the $\phi$ field after
false vacuum decay can be understood in the Heisenberg picture
by using the mode function $\overline{{u_{{\bf k}}(t,{\bf x})}}$
that satisfies the field equation for $\phi$ on the
classical background ${\sigma_0}$ \cite{Ref:TSY,Ref:TS}.
As the boundary condition (\ref{eq:boundary}) does not fix the
normalization of ${g_{{\bf k}}(\tau,{\bf x})}$ and the function $\Omega({\bf
x},{\bf y};\tau)$ is
independent of it, we may assume that the mode
function $\overline{{u_{{\bf k}}(t,{\bf x})}}$ satisfies the Klein-Gordon
normalization
(if it is not the case, we can take a linear combination of them to
satisfy the normalization, {\it i.e.,}
$ v_{\bf p}=\int d^n{\bf k}\lambda({\bf p},{\bf k})u_{{\bf k}}$),
\begin{equation}
-i\int d^n{\bf x}\Bigl(
{u_{{\bf k}}(t,{\bf x})} \overline{\dot u_{{\bf k}'}(t,{\bf x})}-\dot u_{{\bf
k}}(t,x)
\overline{u_{{\bf k}'}(t,{\bf x})}\Big)
=\delta({\bf k}-{\bf k}'),
\end{equation}
where a dot denotes the $t$-differentiation. Introducing the
creation and annihilation operators $\hat a_{\bf k}^\dagger$ and
$\hat a_{\bf k}$,
respectively, associated with these mode functions,
the Heisenberg field operator $\hat \phi_H$ may be expanded as
\begin{equation}
 \hat\phi_H=\int d^n{\bf k} \left(\hat a_{\bf k}{u_{{\bf k}}(t,{\bf x})}+
 \hat a_{\bf k}^\dagger\overline{{u_{{\bf k}}(t,{\bf x})}}\right).
\end{equation}
Then the quantum state described by ${\psi}_L[t,\phi(\cdot)]$
 is found to be a ``vacuum'' state, $\big\vert{\psi}\big>$, defined by
\begin{equation}
\hat a_{\bf k} \big\vert{\psi}\big>=0\quad\hbox{\rm for any} \ {\bf k},
\end{equation}
in the Heisenberg picture.
Note that this ``vacuum'' state is non-trivial in general.
If there exists a natural choice of vacuum, say $\big\vert0\big>$,
the state $\big\vert{\psi}\big>$ is a squeezed state over
$\big\vert0\big>$, described by a non-trivial Bogoliubov transformation.
Then it is possible to discuss the particle creation
 by evaluating the Bogoliubov coefficients between the two states.

\section{particle spectrum}
In this section we first briefly review quantum field theory in the
two-dimensional Milne universe \cite{Ref:BD,Ref:diSessa,Ref:Somm,Ref:Gromes}.
Then we present a method to investigate the quantum state of a field
inside a true vacuum bubble. Here we focus on the case of
two-dimensional spacetime. The extension to the four-dimensional case
will be given in section 4.

The Milne universe is the
hyperbolic time slicing of the Minkowski spacetime in the future
light cone (see Fig.2).
Introducing the coordinates $T$ and $\chi$ in the
two-dimensional Milne universe as
\begin{equation}
t=T\cosh\chi,\quad
x=T\sinh \chi,\qquad
(0<T<\infty, \quad -\infty<\chi<\infty)
\label{eq:coordM}
\end{equation}
the line element is written as
\begin{equation}
ds^2=-dT^2+T^2d\chi^2.
\end{equation}
The field equation of a massive scalar field with mass $M$ is
\begin{equation}
\left[{\partial^2\over\partial T^2}
+{1\over T}{\partial\over\partial T}
-{1\over T^2}{\partial^2\over \partial\chi^2}+M^2\right]
\phi(T,\chi)=0.
\label{eq:apndxa}
\end{equation}
Taking the mode expansion,
\begin{equation}
\phi(T,\chi)=\int_{-\infty}^\infty
 dp~\varphi_p(T) {e^{-ip\chi}\over \sqrt{2\pi}},
\end{equation}
we easily find the solution as
\begin{equation}
\varphi_p(T)=c^{(1)}_p \sqrt{\pi\over 4}e^{-\pi p/2}H^{(1)}_{ip}(MT)
    +c^{(2)}_p \sqrt{\pi\over 4}e^{\pi p/2}H^{(2)}_{ip}(MT),
\label{eq:sola}
\end{equation}
where $c^{(1)}_p$ and $c^{(2)}_p$ are constants, and
$H^{(1)}_{\nu}(z)$ and $H^{(2)}_{\nu}(z)$ are the Hankel functions of
the first and second kinds, respectively.
Following the prescription of the second quantization of a field,
we obtain
\begin{equation}
\phi_H(T,\chi)=\int_{\infty}^{\infty}dp
\biggl[\varphi_p(T) {e^{-ip\chi}\over\sqrt{2\pi}}\hat b_p+
      \overline{\varphi_p(T)} {e^{ip\chi}\over\sqrt{2\pi}}
      \hat b^\dagger_p\biggr],
\label{eq:quantizedfield}
\end{equation}
where $\hat b_p$ and $\hat b_p^\dagger$ are the annihilation and
creation operators, respectively, and $|c^{(2)}_p|^2-|c^{(1)}_p|^2=1$.
The following choice of the coefficients have a special meaning.
\begin{equation}
c^{(1)}_p=0
\hspace{0.5cm}
c^{(2)}_p=1.
\end{equation}
Hereafter we regard $\hat b_p$ and $\hat b_p^\dagger$ as those defined
for this choice of the coefficients.
The calculation of the propagator for the vacuum annihilated by $\hat b_p$
leads to the same expression as that for the usual Minkowski vacuum
\cite{Ref:BD,Ref:diSessa},
\begin{equation}
\bigl<0\vert T\phi_H(x^\mu)\phi_H(x'^\mu)\vert 0\bigr>=
{1\over4i}H^{(2)}_0(-M^2(x^\mu-x'^\mu)^2-i\epsilon).
\end{equation}
Thus $H^{(2)}_{ip}(MT)$ is the positive frequency mode function
for the Minkowski vacuum in the Milne universe.

For later convenience, we show this fact by directly relating the mode
functions in the Milne universe and the usual Minkowski mode functions.
By using the integral representation for the Hankel function
\cite{Ref:Magnus}, we find
\begin{equation}
e^{-i\eta\cosh {\cal K}+i\zeta\sinh {\cal K}}=
{1\over2i}\int_{-\infty}^{\infty}dp e^{-i{\cal K} p}
\Biggl({\eta+\zeta\over \eta-\zeta}\Biggr)^{ip/2}
e^{\pi p/2}H^{(2)}_{ip}\bigl((\eta^2-\zeta^2)^{1/2}\bigr),
\end{equation}
which holds under the condition,
\begin{equation}
{\rm Im}(\eta\pm\zeta)<0.
\label{eq:limit}
\end{equation}
Setting $\zeta=Mx$, $\eta=Mt$ and
${\cal K}={\rm arcsinh}(k/M)$ in the above expression yields
\begin{equation}
e^{-i{\omega_k} t +ikx}={1\over 2i}\int_{-\infty}^{\infty}dp
e^{-i{\cal K} p} \Biggl({t+x\over t-x}\Biggr)^{ip/2}e^{\pi p/2}
H^{(2)}_{ip}\bigl(M(t^2-x^2)^{1/2}\bigr),
\label{eq:trfmb}
\end{equation}
where $\omega_k=M\cosh{\cal K}=\sqrt{k^2+M^2}$.
If we use the coordinates $(T,\chi)$ introduced in Eq.(\ref{eq:coordM}),
we have
\begin{equation}
e^{-i{\omega_k} t +ikx}={1\over 2i}\int_{-\infty}^{\infty}dp
e^{-i{\cal K} p} e^{ip\chi}e^{\pi p/2} H^{(2)}_{ip}(MT).
\label{eq:trfmbb}
\end{equation}
The above formula means that $H^{(2)}_{ip}(MT)$ is indeed
the positive frequency mode function for the Minkowski vacuum.
Note that, because of the validity condition (\ref{eq:limit}),
the presence of a small negative imaginary part in $t$ is understood,
which corresponds to the familiar prescription.

The negative frequency functions are also obtained by taking
the complex conjugate of Eq.(\ref{eq:trfmbb}),
\begin{equation}
e^{i{\omega_k} t -ikx}=-{1\over 2i}\int_{-\infty}^{\infty}dp
e^{i{\cal K} p} e^{-ip\chi}e^{-\pi p/2}
H^{(1)}_{ip}(MT),
\label{eq:negative}
\end{equation}
where we have used the relation,
\begin{equation}
\overline{e^{\pi p/2} H^{(2)}_{ip}(MT)}
=e^{-\pi p/2}H^{(1)}_{ip}(MT).
 \label{eq:Hcc}
\end{equation}
In contrast to the positive frequency mode function,
Eq.(\ref{eq:negative}) is valid on the upper half complex $t$-plane.

Now we consider the mode functions describing the quantum state
after bubble nucleation. Following the general discussion given in the
previous section, we first need to solve the Euclidean field equation to
find the Euclidean mode function $g_k(\tau,x)$.
Introducing the $O(2)$-symmetric coordinates ${\xi_E}$ and $\theta$
in the Euclidean region as
\begin{equation}
x={\xi_E}\sin\theta,\quad\tau=-{\xi_E}\cos\theta,
\qquad
(-{\pi\over 2}\leq\theta\leq{\pi\over 2},\quad 0\leq{\xi_E}<\infty)
\label{eq:coordE}
\end{equation}
the equation for the Euclidean mode function is written as
\begin{equation}
\left[{\partial^2\over\partial {\xi_E}^2}
+{1\over{\xi_E}}{\partial\over\partial {\xi_E}}
+{1\over{\xi_E}^2}{\partial^2\over \partial\theta^2}
-m^2({\sigma_0}({\xi_E}^2))\right]
g_k(\tau,x)=0.
\label{eq:eqofg}
\end{equation}
The boundary condition is $g_k(\tau,x)\rightarrow e^{\omega_k\tau-ikx}$
at $\tau\rightarrow-\infty$.

To find the solution, we first consider a solution of
Eq.(\ref{eq:eqofg}) in the form $F_p({\xi_E}) e^{p\theta}$.
Then $F_p({\xi_E})$ must obey
\begin{equation}
\left[{\partial^2\over\partial {\xi_E}^2}
+{1\over{\xi_E}}{\partial\over\partial {\xi_E}}
+{p^2\over{\xi_E}^2}-m^2(({\sigma_0}({\xi_E}^2))\right]
F_p({\xi_E})=0.
\label{eq:eqofF}
\end{equation}
In the limit ${\xi_E}\rightarrow\infty$, we have
$\sigma_0\rightarrow\sigma_F$.
Let $M^2=m^2(\sigma_F)$, then the asymptotic
solution of $F_p({\xi_E})$ which approaches zero
at ${\xi_E}\rightarrow\infty$
is
\begin{equation}
F_p({\xi_E})=K_{ip}(M{\xi_E}),
\label{eq:bcofF}
\end{equation}
where $K_{\nu}(z)$ is the Modified Bessel function.
On the other hand, from Eq.(\ref{eq:negative}) with
$t\rightarrow -i\tau$ $(\tau<0)$, we find
\begin{equation}
e^{{\omega_k}\tau-ikx}=
{1\over\pi}\int_{-\infty}^{\infty}dp
e^{i{\cal K} p}e^{-p{\theta}}K_{ip}(M{\xi_E}).
\label{eq:EMFB}
\end{equation}
Thus the mode function $g_k(\tau,x)$ can be expressed in the form,
\begin{equation}
g_k(\tau,x)={1\over\pi}\int_{-\infty}^{\infty}dp
e^{i{\cal K} p} e^{-p{\theta}} F_p({\xi_E}),
\label{eq:solg}
\end{equation}
here $F_p({\xi_E})$ is the solution to Eq.(\ref{eq:eqofF})
with the condition $F_p({\xi_E})\rightarrow K_{ip}(M{\xi_E})$
at ${\xi_E}\rightarrow\infty$.

Once we obtain the Euclidean mode function, we continue it to the
Lorentzian region by $\tau\rightarrow it$.
This procedure can be readily performed in the region outside the light
cone $(x^2-t^2>0)$,
\begin{equation}
\overline{u_k(t,x)} ={1\over\pi}\int_{-\infty}^\infty
dp e^{i{\cal K} p}\biggl({t+x\over t-x}\biggr)^{-ip/2} F_p(\xi),
\label{eq:modeuo}
\end{equation}
where $\xi=\sqrt{x^2-t^2}$.
In order to obtain the mode function inside the future light cone,
a bit more consideration is necessary.
To keep the regularity of the mode function,
we must perform the analytic continuation at an infinitesimally small
Euclidean time before $\tau=0$, $\tau\rightarrow it-\epsilon
(\epsilon>0)$, or we must understand Eq.(\ref{eq:modeuo}) with
$t$ replaced by $t+i\epsilon$ \cite{Ref:STYYa}.
In terms of $\xi$, this implies $\xi=e^{-i\pi/2}T$ inside the future
light cone. Then the mode function there is obtained as follows.
In a sufficiently small neighborhood of $\xi=0$, we have
 $m^2(\xi)\simeq {\cal M}^2={\rm const.}$. Hence $F_p(\xi)$ can be
expressed as
\begin{eqnarray}
F_p(\xi) & = & A_pK_{ip}({\cal M}\xi)+B_p\left(I_{ip}({\cal M}\xi)
+I_{-ip}({\cal M}\xi)\right)/2
\nonumber \\
 & = & \tilde A_p K_{ip}({\cal M}\xi)+B_p I_{ip}({\cal M}\xi),
\label{eq:Fpexp}
\end{eqnarray}
where $\tilde A_p =A_p+\displaystyle{i\over\pi}\sinh\pi pB_p$.
Note that, from the reality of $F_p(\xi_E)$,
the coefficients $A_p$ and $B_p$ are real.
Following the above prescription, the Modified Bessel functions
in the above equation become
\begin{eqnarray}
&&K_{ip}({\cal M} \xi)\rightarrow
K_{ip}({\cal M} e^{-i\pi/2}(t^2-x^2)^{1/2})=
{\pi i\over2}e^{-\pi p/2} H_{ip}^{(1)}({\cal M} T),
\\
&&I_{ip}({\cal M} \xi)\rightarrow
I_{ip}({\cal M} e^{-i\pi/2}(t^2-x^2)^{1/2})=
{1\over2}e^{\pi p/2}
\Bigl(H_{ip}^{(1)}({\cal M} T)+H_{ip}^{(2)}({\cal M} T)\Bigr).
\end{eqnarray}
Therefore we have the following expression for the mode function
inside the future light cone at $T\simeq0$,
\begin{equation}
\overline{u_k(t,x)}={1\over\pi}
\int_{-\infty}^\infty dp
e^{i{\cal K} p}e^{-ip\chi}
\biggl[
\overline{c_p}e^{-\pi p/2}H^{(1)}_{ip}({\cal M} T)+
\overline{d_p}e^{\pi p/2}H_{ip}^{(2)}({\cal M} T)\biggr],
\label{eq:modeun}
\end{equation}
where
\begin{eqnarray}
&&c_p:={\pi\over2i}A_p +{\cosh\pi p\over2} B_p,
\nonumber
\\
&&d_p:={1\over2} B_p.
\label{eq:dp}
\end{eqnarray}
Taking the complex conjugate of Eq.(\ref{eq:modeun}), we obtain
\begin{equation}
u_k(t,x)={1\over \pi}
\int_{-\infty}^\infty dp
e^{-i{\cal K} p}
e^{ip\chi}
\biggl[c_{p} e^{\pi p/2}H^{(2)}_{ip}({\cal M} T)+
       d_{p} e^{-\pi p/2}H_{ip}^{(1)}({\cal M} T)\biggr],
\label{eq:modeu}
\end{equation}
where we have used Eq. (\ref{eq:Hcc}).

If we take the limit of no $\sigma$-dependence of mass,
{\it i.e.,} $m^2({\sigma_0}({\xi_E}))=M^2={\rm constant}$ for all ${\xi_E}$,
we have $c_{p}=\pi/2i$ and $d_{p}=0$, and Eq.(\ref{eq:modeu})
reduces to
\begin{equation}
u_k(t,x)={1\over 2i}
\int_{\infty}^\infty dp
e^{-i{\cal K} p}e^{ip\chi}e^{\pi p/2}H^{(2)}_{ip}(MT)
=e^{-i{\omega_k} t+ikx}.
\end{equation}
Thus the particle creation does not occur in this limit,
as expected.

In general, when the width of bubble wall cannot be neglected,
the evolution of the classical background field inside the bubble
becomes important.
This effect leads to time variation of the mass of the
$\phi$ field, and additional particle creation may occur.
In such a case, Eq.(\ref{eq:modeu}) merely gives the (asymptotic) initial
condition of $u_k(t,x)$ at $T\rightarrow0$.

We now focus on the region inside the true vacuum bubble.
Expanding the mode function as
\begin{equation}
u_k(t,x)={1\over\pi}
\int_{-\infty}^\infty dp
e^{-i{\cal K} p}e^{ip\chi}\varphi_p(T),
\label{eq:modeuex}
\end{equation}
the field equation reduces to
\begin{equation}
\left[{\partial^2\over\partial T^2}
+{1\over T}{\partial\over\partial T}
+{p^2\over T^2}+m^2({\sigma_0}(T^2))\right]
\varphi_p(T)=0,
\label{eq:eqofvp}
\end{equation}
thanks to the symmetry of the system. The initial condition
for this equation is specified by Eq.(\ref{eq:modeu}).
We assume that $m^2({\sigma_0}(T^2))\rightarrow\mu^2={\rm constant}$
at $T\rightarrow\infty$, that is, $\sigma_0(T^2)$ settles down to a
finite value (presumably that corresponds to the true vacuum value
 $\sigma_T$).
Then the asymptotic solution at $T\rightarrow\infty$ takes the form,
\begin{equation}
u_k(t,x)={1\over\pi}
\int_{-\infty}^\infty dp
e^{-i{\cal K} p}e^{ip\chi}
\biggl[\tilde c_{p} e^{ \pi p/2}H^{(2)}_{ip}(\mu T)+
       \tilde d_{p} e^{-\pi p/2}H_{ip}^{(1)}(\mu T)\biggr],
\label{eq:modeuasym}
\end{equation}
where $\tilde c_p$ and $\tilde d_p$ depend on the evolutionary history
of $m^2\bigl(\sigma_0(T^2)\bigr)$,

In order to interpret the quantum state described by
 the mode functions obtained in Eq.(\ref{eq:modeuasym}) in the Heisenberg
picture, we need to orthonormalize them.
This is accomplished by taking the following linear combination of ${u_{{\bf
k}}(t,{\bf x})}$:
\begin{equation}
v_q(T,\chi)=\int_{-\infty}^\infty dk \lambda(q,k) u_k(t,x),
\end{equation}
where
\begin{equation}
\lambda(q,k)={e^{i{\cal K} q}\over M \cosh {\cal K}}
{1\over 4\sqrt{2\bigl(
\vert \tilde c_{q}\vert^2-\vert \tilde d_{q}\vert^2\bigr)}}.
\label{eq:lambdapk}
\end{equation}
Then we have
\begin{eqnarray}
v_q(T,\chi)
&&=\biggl[
\alpha_q {\sqrt{\pi}\over 2}e^{\pi q/2}H^{(2)}_{iq}(\mu T)+
\beta_q {\sqrt{\pi}\over 2}e^{-\pi q/2}H^{(1)}_{iq}(\mu T)\biggr]
{e^{iq\chi}\over \sqrt{2\pi}}
\nonumber\\
&&=:\tilde\varphi_q(T)
{e^{iq\chi}\over \sqrt{2\pi}},
\label{eq:normal}
\end{eqnarray}
where
\begin{equation}
\alpha_q:={\tilde c_{q}\over
\sqrt{\vert \tilde c_{q}\vert^2-\vert \tilde d_{q}\vert^2}},
\hspace{0.5cm}
\beta_q:={ \tilde d_{q}\over
\sqrt{\vert \tilde c_{q}\vert^2-\vert \tilde d_{q}\vert^2}},
\label{eq:Bogol}
\end{equation}
or
\begin{equation}
\tilde\varphi_q(T)={\sqrt{\pi}\over 2
\sqrt{\vert \tilde c_{q}\vert^2-\vert \tilde d_{q}\vert^2}}
\varphi_q(T).
\label{eq:normphi}
\end{equation}
One can easily check these mode functions satisfy
 the Klein-Gordon normalization on the $T={\rm constant}$ hypersurface
 in the Milne universe \cite{Ref:Footnote}.

The coefficients $\alpha_q$ and $\beta_q$ are the Bogoliubov
coefficients with respect to the natural vacuum in the Milne universe
which corresponds to the usual Minkowski vacuum.
The spectrum of created particles with respect to
this natural vacuum is
\begin{equation}
n_q=\Big\vert\beta_q\Big\vert^2={1\over
\Bigl\vert {\tilde c_{q}/ \tilde d_{q}} \Bigr\vert^2-1}.
\label{eq:spec}
\end{equation}
When the particle creation after the bubble nucleation can
be neglected, as is the case in the thin wall approximation,
we have $\tilde c_q=c_q$ and $\tilde d_q=d_q$.
Then the spectrum of the particle can be expressed as
\begin{equation}
n_q={1\over \Bigl({\pi A_q/ B_q}\Bigr)^2+\sinh^2\pi q}
   =\left\vert{B_q\over \pi\tilde A_q}\right\vert^2.
\end{equation}

As clear from Eq.(\ref{eq:normphi}), once we have the expression for
$\varphi_p(T)$ in the bubble, it is trivially easy to obtain the
orthonormalized mode functions $v_p(T,\chi)$.
 In other words, the flat Minkowski
coordinates play only an auxiliary role and we may focus on the
evolution of a single mode function whose mode indices are specified with
respect to $(\xi_E,\theta)$- or $(T,\chi)$-coordinates.

Let us summarize our result. First we solve Eq.(\ref{eq:eqofF}) to find
$F_p({\xi_E})$ with the boundary condition that it decreases
exponentially as ${\xi_E}\rightarrow\infty$.
Then we analytically continue it to the interior of the future
light cone $F_p ({\xi_E})\rightarrow\overline{\varphi_p(T)}$ by
${\xi_E}\rightarrow e^{-i\pi/2}T$ and take its complex conjugate.
The resulting function $\varphi_p(T)$ at $T\rightarrow0$ gives the initial
condition of the mode function which
describes the quantum state inside the bubble.
If the evolution of the classical background field $\sigma_0(T^2)$
can be neglected so that the mass of $\phi$
is constant $(={\cal M})$ inside the bubble, as in the
case of a thin-wall bubble, $\varphi_p(T)$ at $T\rightarrow0$
is the only information we need.
By decomposing it into a linear combination of
the asymptotic forms of $e^{\pi p/2}H_{ip}^{(2)}({\cal M} T)$ and
$e^{-\pi p/2}H_{ip}^{(1)}({\cal M} T)$ at $T\rightarrow0$,
the coefficients in front of them
give the (unnormalized) Bogoliubov coefficients with respect to the natural
vacuum of the Milne universe. The normalized Bogoliubov coefficients are
then readily obtained as given in Eq.(\ref{eq:Bogol}).
If the evolution of the classical background cannot be neglected, as in
the case of a thick-wall bubble, we must further solve the equation for
the mode function with the initial condition specified by $\varphi_p(T)$
at $T\rightarrow0$.
Then at $T\rightarrow\infty$ where the mass of $\phi$ settles down to a
constant value ($=\mu$), we again decompose $\varphi_p(T)$ into
the two independent solutions $e^{\pi p/2}H_{ip}^{(2)}(\mu T)$
and $e^{-\pi p/2}H_{ip}^{(1)}(\mu T)$ and read off the
Bogoliubov coefficients with respect to the natural vacuum.
We may then interpret the final quantum state in the language of
 particle creation due to bubble nucleation by considering the particle
spectrum described by the Bogoliubov coefficients.

\section{Extension to four dimension}

The analysis given in the last section can be extended to
a four-dimensional system. We introduce the coordinates
$(T,\chi)$ in the four-dimensional Milne universe as
\begin{equation}
r=T \sinh \chi, \quad t=T \cosh \chi,\qquad
(0<T<\infty, \quad 0<\chi<\infty)
\label{eq:coordMfour}
\end{equation}
where $r$ and $t$ are the radial and time coordinates in the Minkowski
spacetime, respectively.
The line element of the four-dimensional Milne universe is written as
\begin{equation}
ds^2=-dT^2+T^2(d\chi^2+\sinh^2\chi d\Omega^2_{(2)}),
\end{equation}
where $d\Omega^2_{(2)}$ is the line element of the unit two-sphere.
The field equation for a massive scalar field is
\begin{equation}
\left[{\partial^2\over\partial T^2}
+{3\over T}{\partial\over\partial T}
-{1\over T^2} {\bf L}^2
+m^2\right]
\phi(T,\chi,\Omega)=0,
\label{eq:apndxaa}
\end{equation}
where ${\bf L}^2$ is the Laplacian operator on the unit
three-dimensional spatial hyperboloid,
\begin{equation}
{\bf L}^2={1\over \sinh^2\chi}{\partial\over\partial\chi}
\Biggl(\sinh^2\chi{\partial\over\partial\chi}\Biggr)+
{1\over \sinh^2\chi} {\bf L}^2_{\Omega},
\end{equation}
and ${{\bf L}^2_{\Omega}}$ is the Laplacian on the unit two-sphere.
The eigenfunctions $Y_{plm}$ of ${\bf L}^2$
satisfy the equation \cite{Ref:Gromes},
\begin{equation}
{\bf L^2} Y_{plm}(\chi,\Omega)=-(1+p^2)Y_{plm}(\chi,\Omega),
\end{equation}
and is in the form,
\begin{equation}
Y_{plm}(\chi,\Omega)=f_{pl}(\chi)Y_{lm}(\Omega),
\end{equation}
where
\begin{equation}
f_{pl}(\chi)={\Gamma(ip+l+1)\over\Gamma(ip)}{1\over\sqrt{\sinh\chi}}
P^{-l-1/2}_{ip-1/2}(\cosh\chi),
\label{eq:fpl}
\end{equation}
and $Y_{lm}(\Omega)$ is the spherical harmonics on the unit sphere,
$\Gamma(z)$ is the Gamma function and $P^{-l-1/2}_{ip-1/2}(z)$ is
the associated Legendre function of the first kind.
The eigenfunctions $Y_{plm}$ with $0\leq p<\infty$
form a complete orthonormal set for square-integrable functions
on the unit hyperboloid.
Expanding the field operator in terms of $Y_{plm}$, we have
\begin{equation}
\hat\phi_H(T,\chi,\Omega)=\int_0^\infty dp \sum_{lm}
\Biggl[ {\sqrt{\pi}\over 2 } e^{\pi p/2} {H_{ip}^{(2)}(MT)\over T}
Y_{plm}(\chi,\Omega)~\hat b_{plm} + h.c. \ \Biggr].
\label{eq:quantizedfieldb}
\end{equation}
As shown in Appendix, the mode function $e^{\pi p/2} {H_{ip}^{(2)}(MT)/T}$
turns out to be the positive frequency function with respect to the
usual Minkowski vacuum, just the same as in the case of two dimension.

Then we can repeat the procedure given in the previous section to
find the mode function after bubble nucleation.
Introducing the $O(4)$-symmetric coordinates $({\xi_E},\theta)$ in
 the Euclidean region,
\begin{equation}
r={\xi_E}\sin\theta,\quad \tau=-{\xi_E}\cos\theta,
\qquad(0\leq\theta\leq{\pi\over 2},\quad 0\leq{\xi_E}<\infty)
\label{eq:coordEfour}
\end{equation}
the equation for the Euclidean mode functions is written as
\begin{equation}
\left[{\partial^2\over\partial {\xi_E}^2}
+{3\over{\xi_E}}{\partial\over\partial {\xi_E}}
+{1\over{\xi_E}^2}{\bf L}^2_E
-m^2({\sigma_0}({\xi_E}^2))\right]
g_{\bf k}(\tau,{\bf x})=0,
\end{equation}
where
\begin{equation}
{\bf L}^2_E={1\over \sin^2\theta}{\partial\over\partial\theta}
\Biggl(\sin^2\theta{\partial\over\partial\theta}\Biggr)+
{1\over \sin^2\theta} {\bf L}^2_{\Omega}.
\end{equation}
The boundary condition is $g_{\bf k}(\tau,{\bf x})\rightarrow
e^{\omega_k\tau}j_l(kr)Y_{lm}(\Omega)$ at $\tau\rightarrow-\infty$.

Using the formula (\ref{eq:kfourE}) in Appendix, $g_{\bf k}(\tau,{\bf x})$
is expressed as
\begin{eqnarray}
g_{\bf k}(\tau,{\bf x})
=&&(-1)^{l}\sqrt{{2\over\pi}}\int_{0}^\infty p~dp
{P^{-l-1/2}_{-ip-1/2}(\cosh {\cal K})\over \sqrt{\sinh{\cal K}}}
\nonumber
\\
&& \ \ \ \
\times
\sin^l\theta\biggl( {d\over d\cos\theta} \biggr)^{l}
{\sinh p\theta\over \sin\theta}
{F_p({\xi_E})\over M{\xi_E}} ~Y_{lm}(\Omega),
\label{eq:gfour}
\end{eqnarray}
where $F_p({\xi_E})$ satisfies the same equation and the boundary
condition as in the two-dimensional case, Eqs.(\ref{eq:eqofF})
and (\ref{eq:bcofF}). This fact leads to the following expression
for the mode function inside the future light cone,
\begin{eqnarray}
u_{\bf k}(t,{\bf x})=&&(-i)^{l+1} \int_{0}^\infty dp
{\big\vert\Gamma(ip+l+1)\big\vert^2
\over\big\vert\Gamma(ip)\big\vert^2}
{P^{-l-1/2}_{ip-1/2}(\cosh {\cal K})\over \sqrt{\sinh{\cal K}}}
{P^{-l-1/2}_{ip-1/2}(\cosh \chi )\over \sqrt{\sinh\chi }}
\nonumber\\
&& \hspace{1.0cm}
\times
{1\over MT}
\biggl[c_{p} e^{\pi p/2}H^{(2)}_{ip}({\cal M} T)+
       d_{p} e^{-\pi p/2}H_{ip}^{(1)}({\cal M} T)\biggr]
Y_{lm}(\Omega),
\label{eq:MUfour}
\end{eqnarray}
where $c_p$ and $d_p$ are the same as those defined in Eq.(\ref{eq:dp}).
Thus the argument goes completely parallel to the two-dimensional case.

If the background field evolves inside the bubble,
we must further solve Eq.(\ref{eq:eqofvp}) for
the mode function with the initial condition at $T\rightarrow0$
given by Eq.(\ref{eq:MUfour}). The result will take the form,
\begin{eqnarray}
u_{\bf k}(t,{\bf x})=&&(-i)^{l+1}\int_{0}^\infty dp
{\big\vert\Gamma(ip+l+1)\big\vert^2
\over\big\vert\Gamma(ip)\big\vert^2}
{P^{-l-1/2}_{ip-1/2}(\cosh {\cal K})\over \sqrt{\sinh{\cal K}}}
{P^{-l-1/2}_{ip-1/2}(\cosh \chi )\over \sqrt{\sinh\chi }}
\nonumber\\
&& \hspace{1.0cm}
\times
{1\over MT}
\biggl[\tilde c_{p} e^{\pi p/2}H^{(2)}_{ip}(\mu T)+
       \tilde d_{p} e^{-\pi p/2}H_{ip}^{(1)}(\mu T)\biggr]
Y_{lm}(\Omega),
\label{eq:MUfourtil}
\end{eqnarray}
where we have assumed $m^2({\sigma_0}(T^2))\rightarrow\mu^2={\rm constant}$
at $T\rightarrow\infty$.

To obtain the orthonormalized mode functions,
instead of Eq.(\ref{eq:lambdapk}), we set $\lambda(q,k)$ as
\begin{equation}
\lambda(q,k)={\sqrt{\pi}\over2}{i^{l+1}\over\sqrt{
(\vert \tilde c_{q}\vert^2-\vert \tilde d_{q}\vert^2)}}
{(\sinh{\cal K})^{3/2}\over \cosh{\cal K}}
{\Gamma(iq+l+1)\over\Gamma(iq)} P^{-l-1/2}_{iq-1/2}(\cosh{\cal K}),
\end{equation}
and take the linear combination,
\begin{equation}
v_q(T,\chi,\Omega)=\int_0^\infty dk \lambda(q,k) u_{\bf k}(t,{\bf x}).
\end{equation}
Using the orthonormality condition of the harmonic
functions,
\begin{equation}
\int_0^\infty d\chi\sinh\chi
P^{-l-1/2}_{ip-1/2}(\cosh\chi) P^{-l-1/2}_{ip'-1/2}(\cosh\chi)
={\big\vert\Gamma(ip)\big\vert^2\over\big\vert\Gamma(ip+l+1)\big\vert^2}
\delta(p-p'),
\end{equation}
we then find the mode functions which satisfy the Klein-Gordon
normalization
on the $T={\rm constant}$ hypersurface in the Milne universe as
\begin{equation}
v_q(T,\chi,\Omega)=
\biggl(\alpha_q {\sqrt{\pi}\over2}e^{\pi q/2}
{H^{(2)}_{iq}(\mu T)\over T}+
       \beta_q  {\sqrt{\pi}\over2}e^{-\pi q/2}
{H_{iq}^{(1)}(\mu T)\over T}\biggr)
f_{ql}(\chi)
Y_{lm}(\Omega),
\end{equation}
where the coefficients $\alpha_q$ and $\beta_q$ are
 the same as those given by Eq.(\ref{eq:Bogol}).
Thus the particle spectrum is also the same form as in the
two-dimensional case, Eq.(\ref{eq:spec}).

\section{Simple model}

In this section, applying the technique developed in the previous
sections, we consider a simple model of the particle creation
through the tunneling process of bubble nucleation.
We assume a thin-wall bubble so that the mass of the $\phi$ field is
given by
\[m^2\bigl({\sigma_0}({\xi_E}^2)\bigr)=\left\{
\begin{array}{ll}
M^2;\quad &{R}<{\xi_E}<\infty,\\
\mu^2;\quad &0<{\xi_E}<{R},\\
\end{array}\right.\]
where ${R}$ is the radius of the bubble when it nucleates.

The solution to Eq.(\ref{eq:eqofF}) can be easily found
by matching the solutions in both regions,
\[F_p({\xi_E})=\left\{
\begin{array}{ll}
K_{ip}(M{\xi_E});\quad&{R}<{\xi_E}<\infty,
\\
\tilde A_p K_{ip}(\mu{\xi_E})+B_P I_{ip}(\mu{\xi_E});\quad&0<{\xi_E}<{R},
\\
\end{array}\right.\]
where
\begin{eqnarray}
&&\tilde A_p=
\mu R{I_{ip}}'(\mu R){K_{ip}}(MR)
-MR{I_{ip}}(\mu R){K_{ip}}'(MR),
\\
&&B_p=
-\mu R{K_{ip}}'(\mu R){K_{ip}}(MR)
+MR{K_{ip}}(\mu R){K_{ip}}'(MR),
\end{eqnarray}
and a prime denotes differentiation with respect to the
argument.

The particle spectrum is given by
$n_p=\Bigl\vert{B_p}/\pi \tilde A_p\Bigr\vert^2$.
Roughly, the Modified Bessel functions asymptotically behave as
$K_{ip}(z)\propto e^{-\pi p/2}$ and $I_{ip}(z) \propto e^{ \pi p/2}$
for $p\rightarrow\infty$.
Hence the particle spectrum always decreases exponentially for large $p$
as $n_p\propto e^{-2\pi p}$. This suppression of the spectrum at high
momentum limit resembles that of a thermal state.
The same feature has been found in a spatially homogeneous model
of false vacuum decay, first discussed by Rubakov \cite{Ref:Rubakov}
and subsequently by us \cite{Ref:TSY}.
Note that the mode index $p$ corresponds to a comoving wave number
in the Milne universe and the scale $p=1$ corresponds to
the physical curvature scale on the hyperboloid.
In what follows, we evaluate $n_p$ for several limiting cases.

\begin{leftline}
\ (1) $M\ll {R}^{-1}$ and $\mu\ll {R}^{-1}$:
\end{leftline}
\medskip

In this case, we
can use the following asymptotic formulae for the Modified Bessel
functions at $z\ll1$,
\begin{eqnarray}
&&I_{ip}(z) \simeq {(z/2)^{ip}\over \Gamma(ip+1)},
\nonumber
\\
&&K_{ip}(z) \simeq {\pi\over 2}{1\over i\sinh\pi p}
\biggl({(z/2)^{-ip}\over \Gamma(-ip+1)}-
{(z/2)^{ip}\over \Gamma(ip+1)}\biggr).
\label{eq:IKasymsp}
\end{eqnarray}
After some algebra,
we find the following expression for the particle spectrum,
\begin{equation}
n_p\simeq{\sin^2\bigl(p\log(\mu/M)\bigr) \over
\sinh^2 \pi p}.
\end{equation}
We see that $n_p$ decreases exponentially at large $p$,
in accordance with the general discussion in the above.
Due to the sinusoidal dependence, it has zeros in a period of
$\pi/\log(\mu/M)$.
In the limit $p\rightarrow 0$, $n_p$ approaches a finite value,
$n_p\rightarrow\Big\vert\log(\mu/M)/\pi \Big\vert^2$.
Figure 3 shows the spectrum for the case $MR=0.02$ and $\mu R=0.01$.

\begin{leftline}
\ (2) $M\gg{R}^{-1}$ and $\mu\ll {R}^{-1}$:
\end{leftline}
\medskip

In this case, if we focus on the region $p\lesssim1$, the asymptotic formulae
for the Modified Bessel functions at $z\gg1$ can be used,
\begin{equation}
I_{ip}(z) \simeq {1\over \sqrt{2\pi z}} e^{z},\quad
K_{ip}(z) \simeq \sqrt{{\pi\over 2z}} e^{-z},\qquad
{\rm for} \ \ z\gg1.
\label{eq:asymgg}
\end{equation}
Then $n_p$ is evaluated  as
\begin{equation}
n_p={1\over 4\sinh^2\pi p}\Biggl\vert
{\Gamma(ip)\over\Gamma(-ip)}\Bigl({\mu R\over2}\Bigr)^{-2ip}+1
\Biggr\vert^2.
\end{equation}
In the limit $p\rightarrow0$,  we find $n_p\rightarrow
\Big\vert\bigl(\gamma+\log(\mu R/2)\bigr)/\pi \Big\vert^2$, where
$\gamma=0.577\cdots$ is the Euler constant.
Figure 4 shows $n_p$ as a function of $p$ for the case
$MR=20$ and $\mu R=0.01$.

\begin{leftline}
\ (3) $M\gg{R}^{-1}$ and $\mu\gg {R}^{-1}$:
\end{leftline}
\medskip

Using the asymptotic formulae (\ref{eq:asymgg}),
we easily find $n_p$ in the region $p\lesssim1$,
\begin{equation}
n_p\simeq \biggl({\mu-M\over\mu+M}\biggr)^2
e^{-4\mu R}.
\end{equation}

\begin{leftline}
\ (4) $M\ll{R}^{-1}$ and $\mu\gg {R}^{-1}$:
\end{leftline}
\medskip

In this case, using Eqs.(\ref{eq:asymgg}) and (\ref{eq:IKasymsp}),
we obtain $n_p$ at $p\lesssim1$ as
\begin{equation}
n_p\simeq e^{-4\mu R}.
\end{equation}
\bigskip

The first two cases for which $\mu R\ll1$,
the spectrum shows a similar feature, that is,
it is constant of order unity at $p\lesssim1$, and decreases
exponentially at $p\gtrsim1$.
On the other hand, in the latter two cases when $\mu R\gg1$,
the particle creation is always suppressed exponentially by a factor of
$O(e^{-4\mu{R}})$.
This feature is the same as those obtained previously in the
spatially homogeneous decay model of the false vacuum if we identify the
Euclidean tunneling ``time'' in the homogeneous model
with the radius of the bubble wall
\cite{Ref:Rubakov,Ref:Kandrup,Ref:TSY,Ref:Yamamoto}.

\section{Discussion}

In this paper, we have investigated the quantum state of a scalar field
inside a vacuum bubble nucleated in the process of false vacuum decay,
focusing on the aspect of particle creation. We have used the wave functional
formalism \cite{Ref:TSY} to investigate the quantum state.
First, we have carefully analyzed the case of two-dimensional spacetime
and presented a prescription to obtain the mode functions inside a
future light cone of a bubble and an interpretation of the
resulting quantum state in the particle creation picture.
Because of the existence of a natural vacuum in the Milne universe,
which corresponds to the conventional Minkowski vacuum,
we have been able to introduce the concept of ``particle'' in a natural
way. The spectrum of created particles follows immediately by reading
off the Bogoliubov coefficients relative to the
natural vacuum from the final form of the mode functions.
We have shown that the Bogoliubov coefficients are diagonal in the mode
indices so that a mode-by-mode analysis of the quantum state is
possible. Then we have extended our method to four-dimensional spacetime
and shown that it applies in essentially the same manner as in the case
of two dimension.
This technical advantage will be very useful when we consider
realistic models, such as the case of a thick-wall bubble or
the case of quantum fluctuations of the tunneling field itself
\cite{Ref:Hama}.

We have applied the technique developed here to a simple thin-wall model
and found the following.
(1) The spectrum of created particles is always
suppressed exponentially at large wavenumber with the characteristic
scale of the spatial curvature on the hyperbolic time slicing inside the
light cone ({\it i.e.}, the cosmic time slicing in the Milne universe).
(2) If the field becomes massive inside the true vacuum
compared to the inverse of the wall radius, {\it i.e.,}
$\mu \gg {R}^{-1}$, the particle creation is always suppressed
by an exponential factor of $O(e^{-4\mu{R}})$.

In this paper, we have assumed the spacetime to be flat and
only considered a massive field inside the true vacuum bubble.
However, it is of great interest to develop a formalism by relaxing these
restrictions.
In the case of a massive field, the quantum state corresponding
to the Minkowski vacuum state can be written as a pure
state in the Milne universe. But it will not be the case for a massless
field \cite{Ref:pfau}.
This is because the massless modes propagate always along null
directions and a half of them escape to future null infinity
without intersecting the boundary light cone of the Milne universe.
Thus the massless Minkowski mode functions can never be expressed in
terms of those in the Milne universe alone.
Apparently a more careful analysis is necessary for the massless case.

Extension to the case including gravity is most interesting,
especially in connection with the open universe inflation
 \cite{Ref:Recent,Ref:Openinf}.
As pointed out in the Introduction,
if we consider a one-bubble inflationary universe model to realize
a negative curvature universe, the spacetime will be a de Sitter
universe and the quantum fluctuations of the inflation field inside
 a nucleated bubble will give rise to the cosmological density
perturbations.
Hence, if it is possible to include the effect of
gravity in our approach, we will be able to determine the spectrum of
the density perturbations in the open universe inflationary scenario.
This is now under study \cite{Ref:Recent}.



\vskip 0.3in
\centerline{ACKNOWLEDGMENTS}
\vskip 0.05in
We would like to thank Prof. H. Sato for useful comments.
Also we would like to thank B. Allen, R. Caldwell, D. Lyth and N. Turok
for enlightening discussions.
This work was supported in part by Monbusho Grant-in-Aid for
Scientific Research Nos. 2010, 2841 and 05640342, and in part by
the Sumitomo Foundation.

\appendix
\section*{}

In this Appendix, we derive the formulae which relate the mode
functions in the Milne universe and those of the
usual Minkowski spacetime in four-dimensional spacetime.
 As we use the spherical coordinates in both cases,
we omit the spherical harmonics from the mode functions.
We start from the following formula which may be derived from
Eq.(\ref{eq:trfmb}) or (\ref{eq:trfmbb}),
\begin{equation}
k e^{-i{\omega_k} t}j_0(kr)=
-\int_{0}^{\infty}dp~{\sin{\cal K} p }{\sin p\chi\over\sinh\chi}
e^{\pi p/2} {H^{(2)}_{ip}(MT)\over T}.
\label{eq:lzero}
\end{equation}
This gives the relation between the $l=0$ Milne and Minkowski
mode functions.

We extend the above formula to the case of $l\geq1$.
By mathematical induction we can show that
\begin{equation}
\sinh^l\chi \biggl( {d\over d\cosh\chi } \biggr)^l
\biggl(k e^{-i{\omega_k} t}j_0(kr)\biggr)=
{1\over i^l}{\partial^l\over \partial{\omega_k}^l}
\biggl(k^{l+1}e^{-i{\omega_k} t}j_l(kr)\biggr).
\end{equation}
Using this fact,
we operate $(\sinh\chi)^l(d/d\cosh\chi)^l$ on both sides of
Eq.(\ref{eq:lzero})
and integrates both sides of the equation with respect
to ${\omega_k}$, we find
\begin{equation}
k^{l+1}e^{-i\omega_k t}j_l(kr)=-i^l\int_{0}^\infty dp~a_{pk}\,
\sinh^l\chi \biggl( {d\over d\cosh\chi } \biggr)^l
{\sin p\chi\over\sinh\chi}\,
e^{\pi p/2}\, {H^{(2)}_{ip}(MT)\over T},
\end{equation}
where
\begin{eqnarray}
a_{pk}
&=&
\int_{M}^{\omega_k}  d\omega_1
\int_{M}^{\omega_1} d\omega_2
\int_{M}^{\omega_2} d\omega_3 \cdots
\int_{M}^{\omega_{l-1}} d\omega_l
\sin p{\cal K}(\omega_l)
\nonumber\\
&=&
\int_{M}^{\omega_k}  d\omega_2(\omega_k-\omega_2)
\int_{M}^{\omega_2} d\omega_3 \cdots
\int_{M}^{\omega_{l-1}} d\omega_l
\sin p{\cal K}(\omega_l)
\nonumber\\
&=&
\int_{M}^{\omega_k}  d\omega_1{({\omega_k}-\omega_1)^{l-1}\over (l-1)!}
\sin p{\cal K}(\omega_1).
\end{eqnarray}
Noting that $\omega_k=M \cosh {\cal K}$, if we change the integration
variable $\omega_1$ to $\hbox{\rm arccosh} (\omega_1/M)$, we
see that this is just an
integral representation of the associated Legendre function
\cite{Ref:Magnus},
\begin{equation}
a_{pk}=M^l\sqrt{{\pi\over2}} p\,(\sinh {\cal K})^{l+1/2}
P^{-l-1/2}_{ip-1/2}(\cosh {\cal K}).
\end{equation}
Hence we have
\begin{eqnarray}
e^{-i{\omega_k} t}j_l(kr)
= -i^l\int_{0}^\infty && p~ dp \sqrt{{\pi\over2}}\,
{P^{-l-1/2}_{ip-1/2}(\cosh {\cal K})\over \sqrt{\sinh{\cal K}}}\,
\nonumber\\
&&
\times
\sinh^l\chi \biggl( {d\over d\cosh\chi } \biggr)^l
{\sin p\chi\over\sinh\chi}\, e^{\pi p/2}\, {H^{(2)}_{ip}(MT)\over MT}.
\label{eq:kfour}
\end{eqnarray}
Using the formula,
\begin{equation}
\sinh^l\chi \biggl( {d\over d\cosh\chi } \biggr)^l
{\sin p\chi\over\sinh\chi}=
{(-1)^l\over p}\sqrt{{\pi\over2}}
{\big\vert\Gamma(ip+l+1)\big\vert^2
\over\big\vert\Gamma(ip)\big\vert^2}
{P^{-l-1/2}_{ip-1/2}(\cosh\chi)\over \sqrt{\sinh\chi}},
\end{equation}
which can be derived by using the differential relation for
the associated Legendre functions,
we find
\begin{eqnarray}
e^{-i{\omega_k} t}j_l(kr)
= -(-i)^l{\pi\over2} \int_{0}^\infty && dp ~
{\big\vert\Gamma(ip+l+1)\big\vert^2
 \over\big\vert\Gamma(ip)\big\vert^2}
{P^{-l-1/2}_{ip-1/2}(\cosh {\cal K})\over \sqrt{\sinh{\cal K}}}
\nonumber\\
&&
\times
{P^{-l-1/2}_{ip-1/2}(\cosh \chi )\over \sqrt{\sinh\chi }}
e^{\pi p/2} {H^{(2)}_{ip}(MT)\over MT}.
\end{eqnarray}
This result shows that $e^{\pi p/2} {H^{(2)}_{ip}(MT)/ T}$ is indeed
the positive frequency function in the Milne universe which describes
the Minkowski vacuum, as in the two-dimensional case.

By taking the complex conjugate of Eq.(\ref{eq:kfour}), and
continuing it to the Euclidean time $t\rightarrow-i\tau$ $(\tau<0)$,
we obtain
\begin{eqnarray}
e^{{\omega_k}\tau}j_l(kr)
= (-1)^{l}\sqrt{{2\over\pi}}\int_{0}^\infty && p~dp\,
{P^{-l-1/2}_{-ip-1/2}(\cosh {\cal K})\over \sqrt{\sinh{\cal K}}}
\nonumber
\\
&&
\times
\sin^l\theta\biggl( {d\over d\cos\theta} \biggr)^{l}
{\sinh p\theta \over \sin\theta}\, {K_{ip}(m{\xi_E})\over m{\xi_E}}.
\label{eq:kfourE}
\end{eqnarray}
This equation corresponds to Eq.(\ref{eq:EMFB}) in
the two-dimensional case and gives the expression for
the Euclidean mode function in the four-dimensional case.


\vskip 0.3in
\centerline{FIGURE CAPTION}
\vskip 0.05in

\newcounter{fignum}
\begin{list}{Fig.\arabic{fignum}.}{\usecounter{fignum}}

\item An illustration of a potential for the tunneling field
 $\sigma$.
\item A sketch of the coordinates describing the Milne universe.
\item The number spectrum of created particles as a function of
 the wave number $p$ for the case $MR=0.02$ and $\mu R=0.01$.
\item The same as Figure 3, but for $MR=20$ and $\mu R=0.01$.
\end{list}


\begin{thebibliography}{99}
\bibitem{Ref:Footnote}
 Here we are implicitly assuming that
the functional space describing the Cauchy data given on a
 $T={\rm constant}$ hypersurface in the Milne
universe and those on a $t={\rm constant}$ hypersurface in the Minkowski
spacetime are identical. For a field whose mass square is always
positive, this is supported by the
fact that no physical information can escape to the future null
infinity. However, this will be violated for a field whose mass square
becomes zero or negative in a certain (Lorentz invariant)
region of spacetime. A simplest example is the case of a massless field.
We defer detailed discussion of such a case to forthcoming papers
\cite{Ref:Recent,Ref:Hama}.
\end{thebibliography}
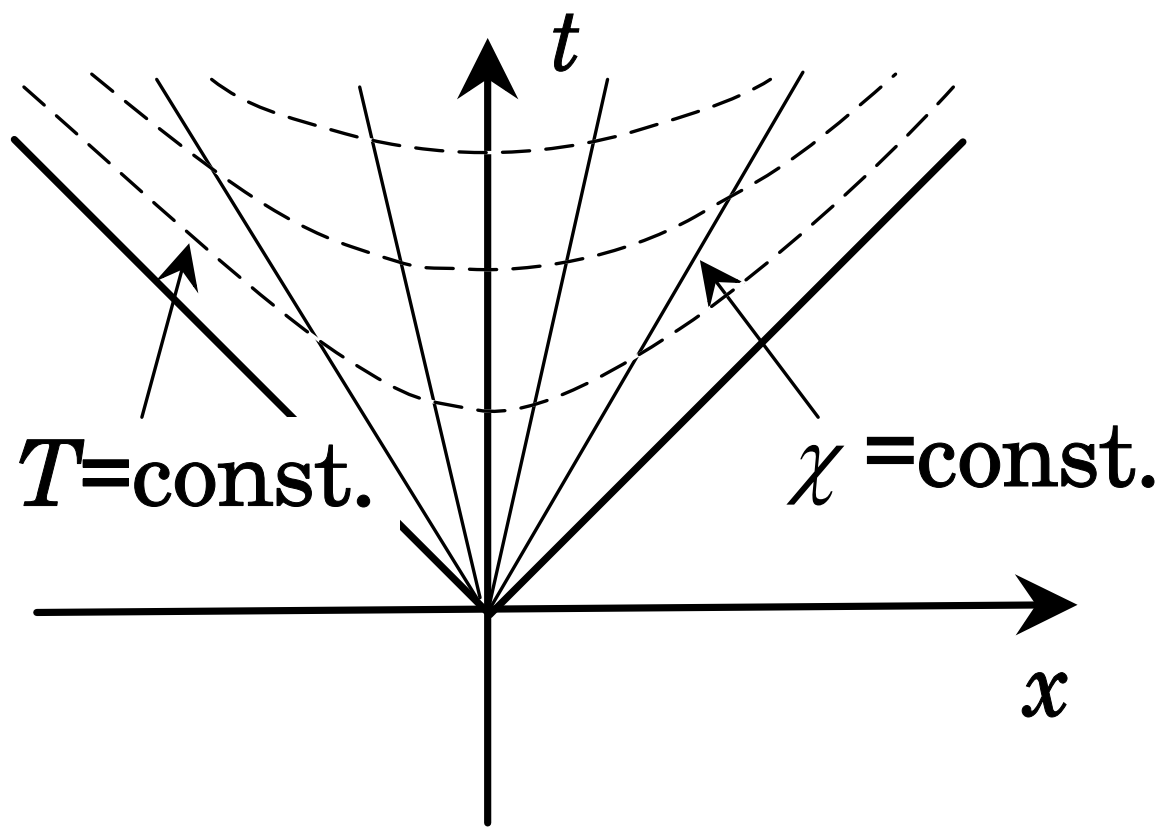
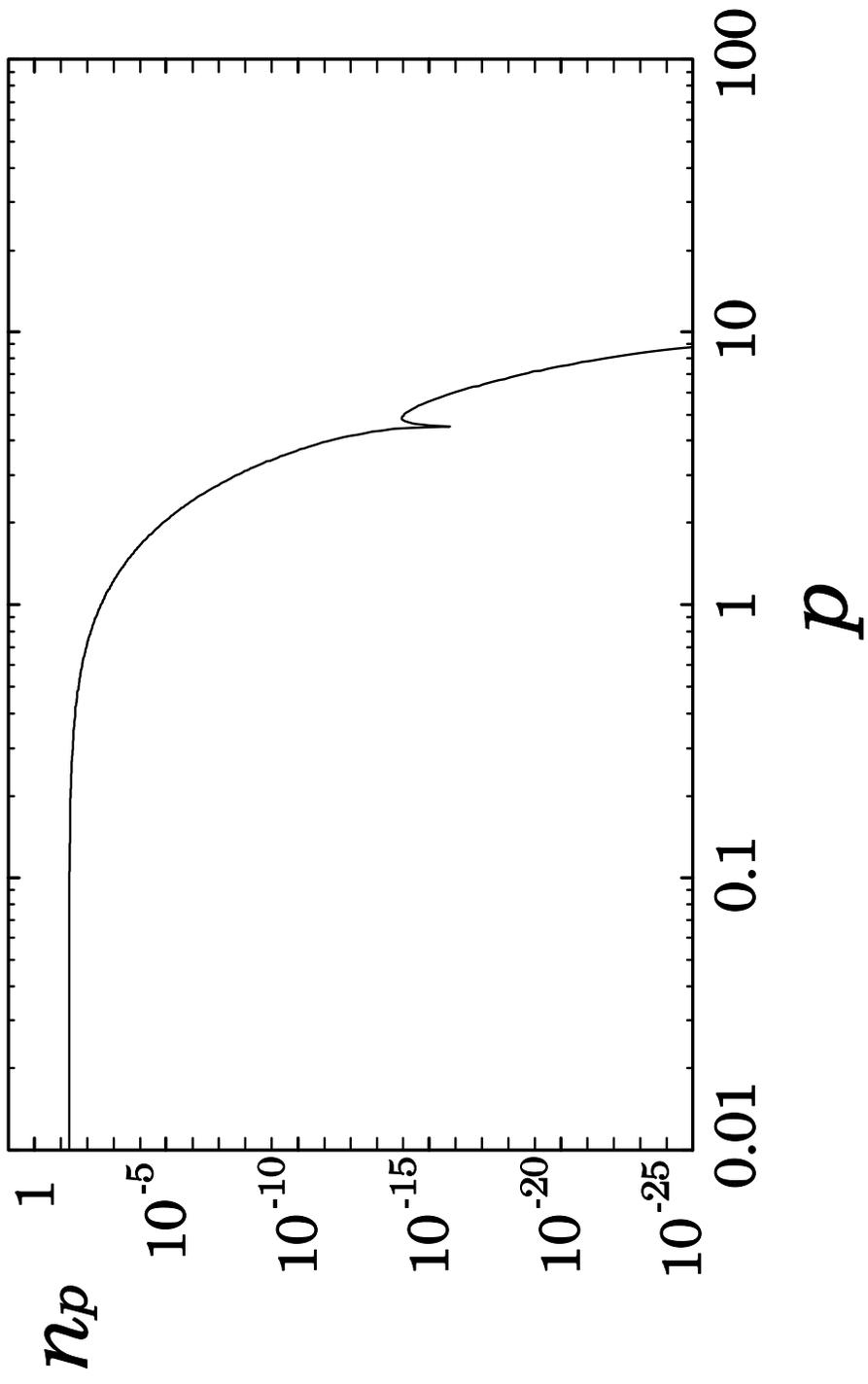
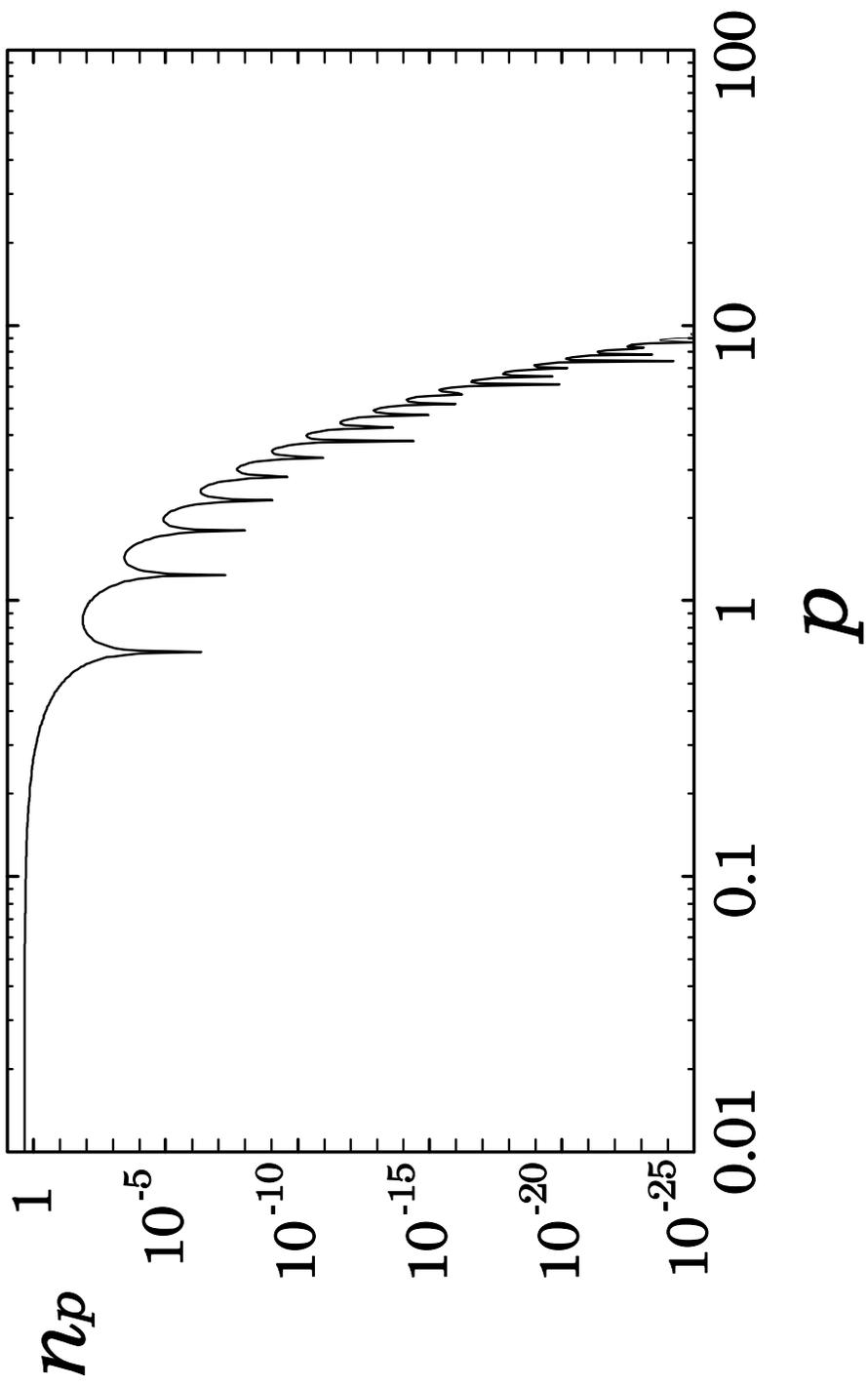
\end{document}